\documentclass[aps,superscriptaddress,prl,floatfix,twocolumn,amsmath,amssymb,english]{revtex4-1}
\usepackage[colorlinks=true, urlcolor=blue, linkcolor=blue, citecolor=blue, pdftex]{hyperref}
\usepackage{graphicx}
\usepackage{latexsym}
\usepackage{amsmath}
\usepackage{wasysym}
\usepackage{nicefrac}
\usepackage{color}
\usepackage{dcolumn}
\usepackage{bm}
\usepackage{footmisc}
\usepackage{braket}

\newcommand{\dagga}{{\phantom{\dagger}}}

\begin{document}

\title{Intertwined superconductivity and orbital selectivity in a three-orbital Hubbard model for the iron pnictides}
\author{Vito Marino}
\affiliation{Institute for Condensed Matter Physics and Complex Systems, DISAT, Politecnico di Torino, I-10129 Torino, Italy}
\affiliation{International School for Advanced Studies (SISSA) and CNR-IOM, Via Bonomea 265, I-34136 Trieste, Italy}
\author{Alberto Scazzola}
\affiliation{Department of Electronics and Telecommunications, Politecnico di Torino, I-10129 Torino, Italy}
\author{Federico Becca}
\affiliation{Dipartimento di Fisica, Universit\`a di Trieste, Strada Costiera 11, I-34151 Trieste, Italy}
\author{Massimo Capone}
\affiliation{International School for Advanced Studies (SISSA) and CNR-IOM, Via Bonomea 265, I-34136 Trieste, Italy}
\author{Luca F. Tocchio}
\affiliation{Institute for Condensed Matter Physics and Complex Systems, DISAT, Politecnico di Torino, I-10129 Torino, Italy}
\date{\today}

\begin{abstract}
We study a three-orbital Hubbard-Kanamori model relevant for iron-based superconductors using variational wave functions explicitly including spatial 
correlations and electron pairing. We span the nonmagnetic sector from filling $n=4$, which is representative of undoped iron-based superconductors, 
to $n=3$, where a Mott insulating state with each orbital at half filling is found. In the strong-coupling regime, when the electron density is increased, 
we find a spontaneous differentiation between the occupation of $d_{xz}$ and $d_{yz}$ orbitals, leading to an orbital-selective state with a nematic 
character that becomes stronger at increasing density. One of these orbitals stays half filled for all densities while the other one hosts (together 
with the $d_{xy}$ orbital) the excess of electron density. Most importantly, in this regime long-range pairing correlations appear in the orbital with 
the largest occupation. Our results highlight a strong link between orbital-selective correlations, nematicity, and superconductivity, which requires 
the presence of a significant Hund's coupling. 
\end{abstract}

\maketitle

{\it Introduction.} 
The discovery of iron-based superconductors (IBS)~\cite{kamihara2008} expanded the realm of unconventional superconductivity, inviting comparisons with 
copper-based high-$T_c$ superconductors (cuprates). In both families, superconductivity emerges in rich phase diagrams and competes with quantum phases driven 
by electron-electron interactions, including spin-density waves and nematic metallic states. 

An undisputed difference is the fact that in IBS the low-energy electronic structure features more than one band, calling for the inclusion of different atomic 
orbitals in the modeling, while copper-based materials are well described in a single-band picture. This entails important consequences for the degree and the 
nature of electronic correlation in these materials~\cite{fernandes2022}. Indeed, conflicting evidences led to a persistent debate as to whether a weak- or a 
strong-coupling description is more appropriate~\cite{haule2008,mazin2008,raghu2008,si2008,wang2009,chubukov2012,fernandes2014}. In particular, a picture in 
terms of renormalized itinerant electrons coupled by bosonic excitations, typically of magnetic origin, accounts for several features of the superconducting 
phase and its interplay with nematic ordering~\cite{chubukov2015,chubukov2016,yao2018,kreisel2017,hu2018,benfatto2018}. On the other hand, the presence of a 
sizable Hund's coupling has been identified as the underlying principle that leads to anomalies of the normal state (the so-called Hund's 
metal)~\cite{haule2008,yin2011,georges2013} among which a pivotal role is played by orbital-selective correlations~\cite{demedici2014}, i.e., by the simultaneous 
presence of strongly correlated and weakly correlated electrons associated with different atomic orbitals. An orbital-selective character has been also reported 
experimentally both for superconductivity~\cite{sprau2017} and the normal state~\cite{yi2015,kostin2018,capone2018}. Indeed, the Hund's metal has been shown 
to favor an enhancement (or even a divergence) of the charge compressibility~\cite{demedici2017,villar2018,chatzieleftheriou2023} and to favor both 
superconductivity~\cite{nica2017,fanfarillo2020} and nematic ordering~\cite{fanfarillo2023}.

\begin{figure*}
\includegraphics[width=\textwidth]{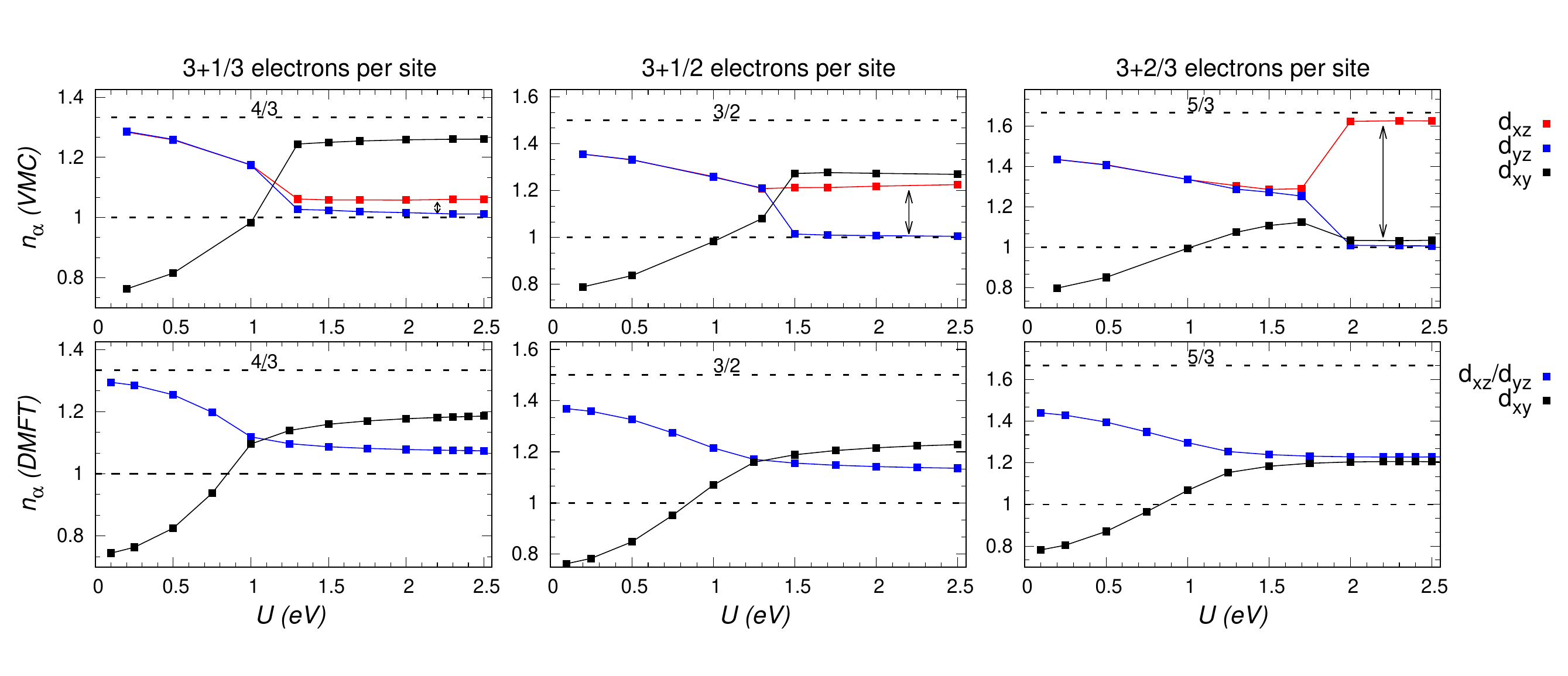}
\caption{\label{fig:density} 
Upper panels: Electronic density per orbital $n_{\alpha}$ as a function of $U$, as obtained within the VMC method for three different total electronic 
densities: $n=3+1/3$ (left column), $n=3+1/2$ (middle column), and $n=3+2/3$ (right column). We fix $J/U=0.2$. Data are for the $L=12\times 12$ cluster. 
The arrows highlight the size of the differentiation between the $d_{xz}$ and the $d_{yz}$ orbitals. Lower panels: The same as in the upper panels, computed 
with DMFT and an exact-diagonalization solver.}
\end{figure*}

In this scenario, where seemingly contradictory approaches have been successfully used, several fundamental questions remain open. A particularly important 
and challenging problem is to understand whether superconductivity requires the explicit inclusion of a boson mediating pairing, or whether superconductivity 
arises directly within the intermediate and/or strong coupling regime of the local interactions. This question is very elusive also for technical reasons. In fact, 
it is very hard to find a methodology that is able to simultaneously capture the effects of strong correlations and the low-energy bosonic mediators for the 
multi-orbital models, which are necessary to properly study these materials. In Ref.~\cite{fanfarillo2020}, a simplified approach to superconductivity in 
Hund's metals (inspired by Ref.~\cite{capone2004,capone2009}) has been implemented, showing that a boson-mediated pairing is compatible with a Hund's metal 
that in turn introduces orbital-selective features. In general, most of the information about the role of Hund's correlations comes from Dynamical Mean-Field 
Theory (DMFT)~\cite{georges1996}, which introduces accurate dynamical local correlations, or simpler methods sharing the local focus such as slave 
particles~\cite{demedici2017book}. However, all these methods do not include non-local correlations beyond the static mean-field approximation.

In this Letter, we make a further step forward by using the variational Monte Carlo (VMC) method to investigate the three-orbital Hubbard-Kanamori 
model~\cite{kanamori1963,georges2013} used in Ref.~\cite{fanfarillo2020,fanfarillo2023} and adapted from Ref.~\cite{daghofer2010,fernandes2017}. The VMC 
method allows us to accurately describe non-local correlations and ordering, including non-local superconductivity, none of them being accessible within 
DMFT. We work in the nonmagnetic sector to emphasize the intrinsic Hund-driven correlation effects and to establish their relation with superconductivity 
and orbital symmetric breaking. We focus on the hole-doped region by changing the filling from $n=4$ which corresponds, in our three-orbital model, to the 
undoped compounds (which have a filling of 6 electrons in 5 orbitals) to $n=3$, where a Mott insulator is found~\cite{demedici2014}. 

\begin{figure*}
\includegraphics[width=\textwidth]{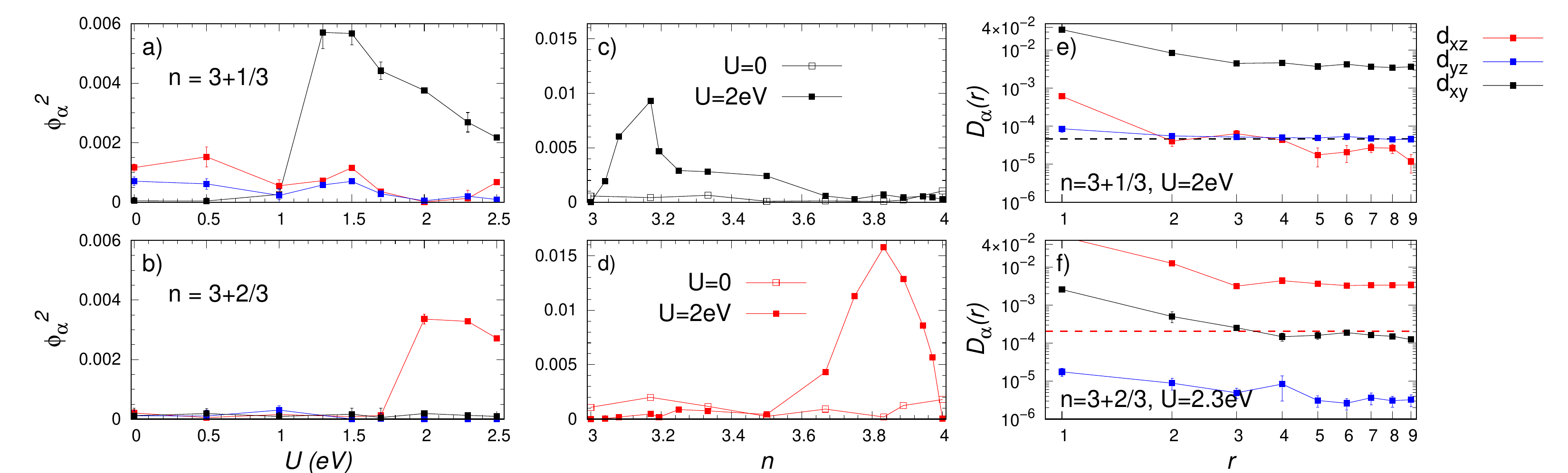}
\caption{\label{fig:Superc}
Square of the superconducting order parameter $\phi_{\alpha}^2$ in each orbital as a function of $U$ at $n=3+1/3$ [panel a)] and $n=3+2/3$ [panel b)]. 
Data are for the $L=18\times 18$ cluster. Square of the superconducting order parameter in the $d_{xy}$ [panel c)] and $d_{xz}$ [panel d)] orbitals 
as a function of $n$ at $U=2$eV (full squares) and at $U=0$ (empty squares). Data are for the $L=12\times 12$ cluster. Pairing correlation functions 
$D_{\alpha}(r)$ in all orbitals, for a value of $U$ within the orbital-selective regime at total electronic density $n=3+1/3$ [panel e)] and $n=3+2/3$ 
[panel f)]. Data are for the $L=18\times 18$ lattice; the dashed lines denote $\phi_{xy}^2$ [panel e)] and $\phi_{xz}^2$ [panel f)] at $U=0$. In all cases $J/U=0.2$.}
\end{figure*}

We observe an ubiquitous spontaneous breaking of the orbital degeneracy between $d_{xz}$ and $d_{yz}$ orbitals that takes place moving away from $n=3$, where 
every orbital is half filled. Such a differentiation implies a nematic order in the metallic state. For large electron-electron repulsion, one of the two 
orbitals remains half filled in the whole density window, while the other becomes more and more filled as we approach $n=4$, suggesting an orbital-selective 
Mott transition~\cite{demedici2009}. Most importantly, the orbital with the largest occupation has sizable pairing correlations, within the orbital selective 
phase. We underline that Hund's coupling is necessary to drive both orbital selectivity and superconductivity.

{\it Model and Method.}
We consider the three-orbital ($d_{xz}$, $d_{yz}$, and $d_{xy}$) Hamiltonian ${\cal H}_0$~\cite{fanfarillo2020,fanfarillo2023} that reproduces qualitatively 
the generic shape and the orbital content of the Fermi surfaces of IBS: Two hole-like pockets around the $\Gamma$ point and two elliptic electron-like pockets 
centered at the symmetry-related points $X$ and $Y$. The explicit form of the Hamiltonian is reported in the Supplemental Material~\cite{supplementary}.
Importantly, the degeneracy of the three orbitals is partially lifted by the crystal field (penalizing the $d_{xy}$ orbital over the $d_{yz}$ and $d_{zx}$).
As in previous works, we include local Hubbard and Hund's couplings~\cite{kanamori1963,georges2013}. We underline that our model is not chosen to reproduce 
detailed features of any individual IBS, but rather as a simplified tight-binding picture, which is sufficiently light to allow for an accurate numerical 
solution, while containing the main features of the IBS fermiology. For this reason, we will not attempt a direct comparison with experiments, but we will 
rather extract information about the general trends arising from the interplay between non-local electronic correlations and a sensible electronic structure.

The VMC approach is based on the definition of correlated variational wave functions, whose parameters and properties can be evaluated within a Monte Carlo 
scheme~\cite{becca2017}. In particular, the electron-electron correlation is inserted by Jastrow factors~\cite{capello2005,defranco2018} on top of an 
uncorrelated Bardeen-Cooper-Schrieffer (BCS) state, which is appropriate to describe superconductivity:
\begin{equation}\label{eq:psi}
\ket{\Psi} = \mathcal{J}_{c} \mathcal{J}_{s} \ket{\Phi_0}.
\end{equation}
Here, $\mathcal{J}_{c}$ and $\mathcal{J}_{s}$ are density and spin Jastrow factors that are suitable to include correlations within the variational state
$\mathcal{J}_{c} = \exp{ \Big( -\frac{1}{2} \sum_{\alpha,\beta} \sum_{R,R^\prime} v_{R,R^\prime}^{\alpha,\beta} n_{R,\alpha} n_{R^\prime,\beta} \Big)}$ 
and $\mathcal{J}_{s} = \exp{ \Big( -\frac{1}{2} \sum_{\alpha \neq \beta} u^{\alpha,\beta} \sum_{R} S_{R,\alpha}^z S_{R,\beta}^z \Big)}$ where 
$n_{R,\alpha}=n_{R,\alpha,\uparrow}+n_{R,\alpha,\downarrow}$ and $S^z_{i,\alpha}=(n_{R,\alpha,\uparrow}-n_{R,\alpha,\downarrow})/2$ are the total density 
and spin along the $z$ axis on site $R$ and orbital $\alpha$, i.e., $n_{R,\alpha,\sigma}= c^\dag_{R,\alpha,\sigma} c^\dagga_{R,\alpha,\sigma}$ (where 
$c^\dag_{R,\alpha,\sigma}$ and $c^\dagga_{R,\alpha,\sigma}$ are creation and annihilation operators for fermions on site $R$, orbital $\alpha$, and spin 
$\sigma$). The parameters $v_{R,R'}^{\alpha,\beta}$ and $u^{\alpha,\beta}$ are optimized to minimize the variational energy~\cite{sorella2005,becca2017}. 
Notice that the density Jastrow factor includes long-range terms, which are important to describe the Mott insulator~\cite{capello2005,defranco2018}, 
while the spin-spin ones are limited to on-site terms, in order to include the Hund's effect. The uncorrelated state $\ket{\Phi_0}$ is the ground state 
of an auxiliary Hamiltonian featuring a BCS intra-orbital pairing, in addition to ${\cal H}_0$:
\begin{equation}
 \begin{split}
& \mathcal{H}_{BCS} = \mathcal{H}_{0} - \sum_{R,\alpha,\sigma} \mu_{\alpha} c_{R,\alpha,\sigma}^\dag c_{R,\alpha,\sigma}  \\
    &+ \sum_{R,\alpha,\delta} \left[ \Delta_{\alpha,\delta} \left( c^\dagga_{R,\alpha,\uparrow} c^\dagga_{R+\delta,\alpha,\downarrow}
     - c^\dagga_{R,\alpha,\downarrow} c^\dagga_{R+\delta,\alpha,\uparrow} \right) + \textrm{h.c.}  \right],
\label{eq:H_aux}
\end{split}
\end{equation}
where $\delta=x$, $y$, $x+y$, and $x-y$ indicates nearest and next-nearest neighbors of the site $R$ and $\mu_{\alpha}$ defines the chemical potential 
of orbital $\alpha$. The (singlet) pairing amplitudes $\Delta_{\alpha,\delta}$ and $\mu_{\alpha}$ are also optimized, while the hopping parameters in 
$\mathcal{H}_{0}$ are kept fixed to the bare Hamiltonian. Inter-orbital pairing amplitudes are found to be negligible in the optimal wave function, in 
agreement with~\cite{misawa2014}. Calculations are done on $L= l \times l$ clusters, with periodic boundary conditions, mainly with $l=12$.

{\it Results.} 
We start with the emergence of the orbital selectivity. In Fig.~\ref{fig:density}, we report the density occupation of the orbitals $\{ n_{xz}, n_{yz}, n_{xy} \}$ 
for a few representative densities $n=3+\delta n$ (with $\delta n=2/3$, $1/2$, and $1/3$) and different values of the Hubbard-$U$ interaction with $J/U=0.2$. 
The results are compared with DMFT using an exact-diagonalization solver~\cite{EDIpack}. For all the cases (including others not shown), the results exhibit the 
same behavior for small values of the interaction strength $U$: the two orbitals $d_{xz}$ and $d_{yz}$ have the same electron occupation, which is larger than 
the one of the orbital $d_{xy}$ (penalized by the crystal field). By increasing $U$, the difference among the orbitals reduces, since the interaction terms 
affect all of them in the same way. This initial trend is also found in DMFT. For even larger values of $U$, the VMC results change drastically. Indeed, for 
$0< \delta n \lesssim 0.5$, the $d_{xy}$ orbital (which has sizable intra-orbital hoppings) becomes the most occupied one, while the other ones have smaller 
occupations, with one of them (e.g., $d_{yz}$) being half filled and the other one ($d_{xz}$) having an intermediate occupation, i.e., 
$n_{xy}>n_{xz}>n_{yz} \approx 1$. This fact implies a nematic feature, which increases moving away from $n=3$, where no sign of orbital selectivity has been 
detected. For $\delta n$ slightly larger than 1/2, the occupations of $d_{xz}$ and $d_{xy}$ orbitals cross each other. By further increasing $\delta n$, 
$n_{xy}$ tends rapidly to half filling ($n_{xy} \to 1$), while most of the excess electron density goes into the $d_{xz}$ one ($n_{xz} \approx 1+\delta n$). 
We remark that the global minimum of the variational energy always gives a differentiation between $d_{xz}$ and $d_{yz}$ orbitals; still, whenever the 
density-density Jastrow factor is limited to on-site terms (i.e., forcing to have $v_{R,R'}^{\alpha,\beta}=0$ for $R \ne R'$), a local minimum with no density
disproportion between $d_{xz}$ and $d_{yz}$ orbitals exists. In addition, nematicity is not obtained within the DMFT approach (which, however, predicts the 
orbital occupation switch). All these facts strongly suggest that nematicity and density disproportion are driven by (long-range) spatial correlations.

Now, we turn to the main result of our work, namely the possible emergence of superconductivity. In order to assess the true presence of a superconducting 
state, we compute the intra-orbital pairing correlations $D_{\alpha}(r) = \frac{1}{L} \sum_{R} \langle P^\dagga_{R,\alpha} P^\dag_{R+rx,\alpha} \rangle$
where $P^\dagga_{R,\alpha} = c^\dagga_{R+y,\alpha,\downarrow} c^\dagga_{R,\alpha,\uparrow} - c^\dagga_{R+y,\alpha,\uparrow} c^\dagga_{R,\alpha,\downarrow}$ 
destroys two electrons on the same orbital at nearest-neighbor sites (along $y$). Then, superconductivity exists in orbital $\alpha$ whenever 
$\phi_{\alpha}^2=\lim_{r \to \infty} D_{\alpha}(r)$ is finite in the thermodynamic limit.

\begin{figure}
\includegraphics[width=\columnwidth]{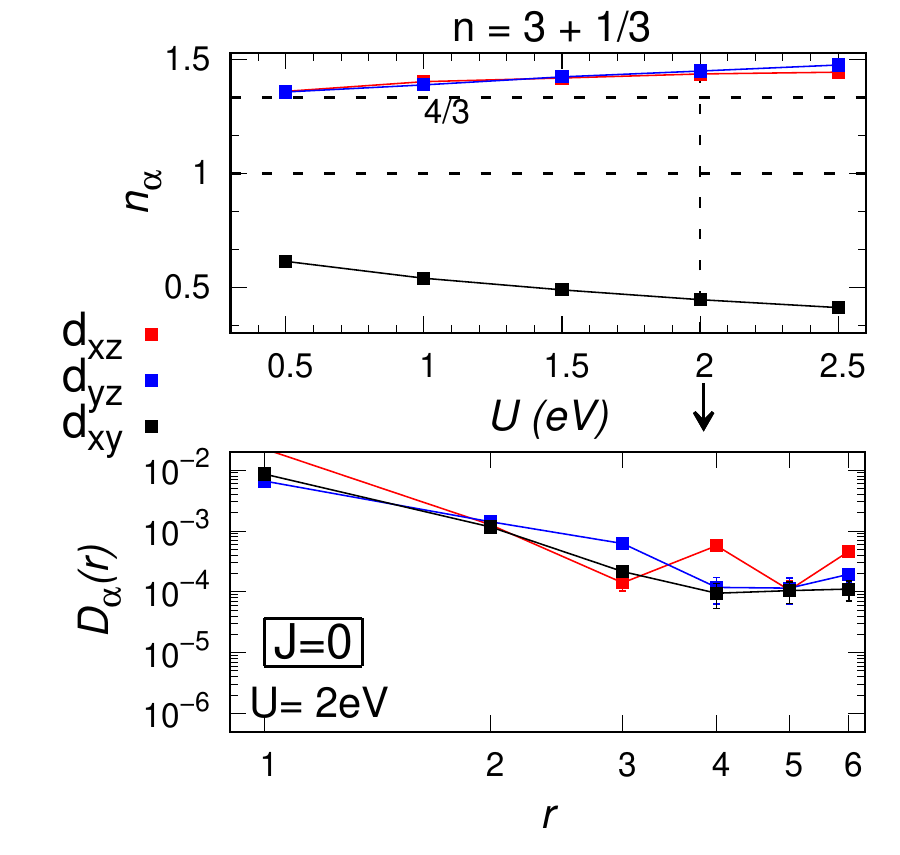}
\caption{\label{fig:NoJ}
Electronic density per orbital $n_{\alpha}$, as a function of $U$ (upper panel) and pairing-pairing correlations $D_{\alpha}(r)$ for $U=2$eV (lower panel), in 
the absence of Hund's coupling at density $n=3+1/3$. Data are for a $L=12 \times 12$ cluster.} 
\end{figure}

\begin{figure}
\includegraphics[width=\columnwidth]{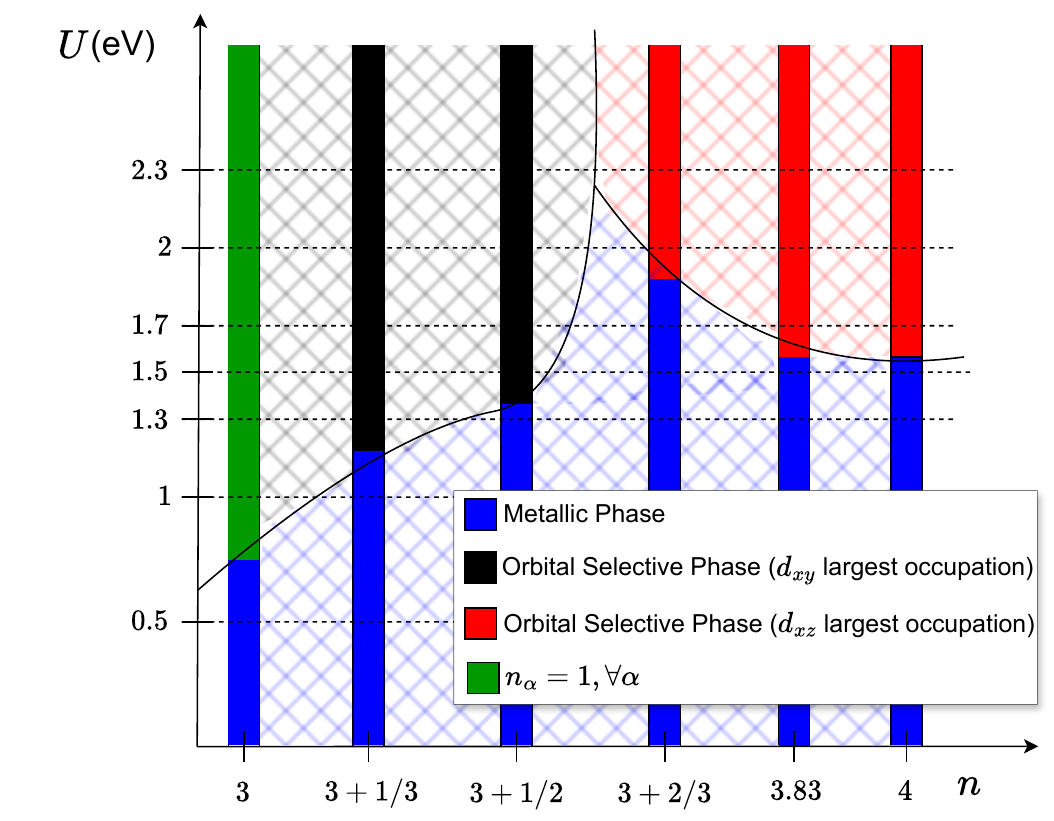}
\caption{\label{fig:PD}
Ground-state phase diagram of the three-orbital model, as reconstructed from the calculations on selected electron densities. The Hund's coupling is fixed to 
$J/U=0.2$. The red and the black regions are characterized by the development of superconductivity, as discussed in the text.} 
\end{figure}

Our results indicate that superconductivity emerges in the most filled orbital, that is away from half filling, within the orbital selective region. The results
for $\phi_{\alpha}^2$ as a function of $U$ are shown in Fig.~\ref{fig:Superc}, for $n=3+1/3$ and $n=3+2/3$. In the former case, superconductivity develops in 
the $d_{xy}$ orbital, inside the orbital selective phase (see Fig.~\ref{fig:density}). The other two orbitals, instead, do not show any enhancement of the 
superconducting correlations with respect to the $U=0$ case (which is finite on finite clusters). In the latter case, superconductivity develops in the $d_{xz}$ 
orbital, again within the orbital selective phase. Remarkably the highest signal in the pairing correlations is detected in the vicinity of the orbital-selective 
transition, while at large values of the electron-electron repulsion pairing is suppressed (as in the one-band Hubbard model~\cite{tocchio2016}). The behavior 
of $\phi_{\alpha}^2$ as a function of density $n$ is also shown in Fig.~\ref{fig:Superc}, for a value of $U$ that lies within the orbital-selective region. 
In order to emphasize the relevance of the Hubbard-$U$ to generate the electron pairing, we also report the non-interacting values (on the same cluster). 
A visible dome-like behavior is present for $3 \lesssim n \lesssim 3.5$ in the pairing correlations of the $d_{xy}$ orbital, with a clear enhancement with 
respect to the noninteracting case. By contrast, pairing correlations are strengthened within the $d_{xz}$ orbital for $3.5 \lesssim n \lesssim 4$. Notice that, at 
integer fillings, where insulators are obtained, pairing correlations are suppressed going from the noninteracting to the interacting case. Finally, the full 
behavior of $D_{\alpha}(r)$ is reported within the orbital selective region, for $n=3+1/3$ and $n=3+2/3$. In both cases, the superconductive orbital shows 
superconductive correlations that are clearly enhanced with respect to the $U=0$ case (that is shown as a horizontal dashed line). A full comparison between 
$D_{\alpha}(r)$ within the orbital selective region and at $U=0$ is discussed in the Supplemental Material~\cite{supplementary}. 

The orbital-selective and superconducting region is characterized by the development of sizable BCS parameters in both $d_{xy}$ and $d_{yz}$ orbitals, while they 
are relatively small in the $d_{xz}$ one~\cite{supplementary}. At $n=3+1/3$, the existence of superconducting correlations in the $d_{xy}$ orbital is accompanied 
by sizable pairing amplitudes of the $d_{xy}$ orbital in Eq.~\eqref{eq:H_aux}. If we Fourier transform the pairing amplitudes $\Delta_{xy,\delta}$ to $k$-space 
in the nematic regime, we find that $\Delta_{xy,k}$ resembles the $s^{\pm}$ symmetry~\cite{mazin2008,chubukov2008,seo2008}, with a strong difference between $X$ 
and $Y$ point that is due to the underlying nematicity~\cite{supplementary}. The situation is different for $n=3+2/3$, where the $d_{xz}$ orbital shows sizable 
pairing correlations, despite having relatively small BCS parameters~\cite{supplementary}. This fact is only possible because of the inter-orbital hybridization 
in the BCS Hamiltonian~\eqref{eq:H_aux}.

In order to highlight the role of the Hund's coupling, we present calculations for $J=0$, showing that two orbitals $d_{xz}$ and $d_{yz}$ remain equally filled 
and more occupied than the $d_{xy}$ one, for all the values of $U$ and no superconducting state is found, see Fig.~\ref{fig:NoJ} for $n=3+1/3$. Therefore, it 
turns out that the symmetry breaking between $d_{xz}$ and $d_{yz}$ orbitals and superconductivity are intimately related (and their common origin requires a 
sizable $J/U$). 

We finally notice that, even for purely local interactions, the symmetry of the order parameter depends on the shape of the Fermi surface. Indeed a simpler 
tight-binding Hamiltonian $\mathcal{H}_{0}$ with the same diagonal nearest-neighbor hopping for every orbital, displays superconductivity for $n=3+1/3$, but 
with the $d_{x^2-y^2}$ symmetry~\cite{supplementary}, that one obtains for the one-band Hubbard model~\cite{tocchio2016}. 

{\it Conclusions.} 
By using variational wave functions including pairing parameters and Jastrow factors, we have studied a simplified model for IBS superconductors. Spanning the 
density range between $n=4$ (parent compound) and $n=3$ (Mott insulator), we find a nematic state with orbital-selective character and superconducting pairing. 
In particular, the $d_{yz}$ (or $d_{xz}$) orbital remains half-filled in the whole density regime, thus leading to a spontaneous orbital-selective Mott transition
and a broken rotational symmetry. Most importantly, superconductivity is observed in the nematic regime, changing its character when the electron doping is varied. 
For $3<n \lesssim 3.5$, the $d_{xy}$ orbital sustains long-range electron pairing. Instead, for $3.5 \lesssim n < 4$, pairing correlations are dominant in the 
$d_{xz}$ (or $d_{yz}$) orbital. All these features are triggered by the presence of Hund's coupling, since neither orbital selectivity nor superconductivity are 
found for $J=0$. The tentative phase diagram for $3 \le n \le 4$ is shown in Fig.~\ref{fig:PD}.

Our work underlines how non-local correlations enrich the scenario of Hund's driven correlations with respect to DMFT, and establish a strong link between nematic 
correlations and superconductivity. The results are reminiscent of several observations in IBS, but we believe that a closer comparison with experiments must be 
deferred to future studies of more involved and complete models.

{\it Acknowledgements.} We acknowledge useful discussions with L. Fanfarillo. Computational resources were provided by HPC@POLITO (http://www.hpc.polito.it). We acknowledge also the 
CINECA award under the ISCRA initiative, for the availability of high performance computing resources and support. M.C. acknowledges financial support of MUR via 
PRIN 2020 (Prot. 2020JLZ52N 002) programs, and by the European Union - NextGenerationEU through PRIN 2022 (Prot. 20228YCYY7), PNRR MUR Projects No. PE0000023-NQSTI, and No. CN00000013-ICSC. The views and opinions expressed are solely those of the authors and do not necessarily 
reflect those of the European Union, nor can the European Union be held responsible for them.

\section{Supplemental Material}

In this Supplemental Material, we show the tight-binding model ${\cal H}_0$ that is used for the three-orbital Hubbard Hamiltonian considered in this work. Then, we present a comparison between the intra-orbital pairing correlations $D_{\alpha}(r)$ at density $n=3+1/3$ and $n=3+2/3$,  within the orbital selective region and at $U=0$. Moreover, we show the optimal BCS parameters as a function of $U$ at density $n=3+1/3$ and $n=3+2/3$, as well as the symmetry of the BCS parameters in $k$-space at density $n=3+1/3$. Finally, we present results for the BCS parameters and for the pairing 
correlations when the tight-binding Hamiltonian is studied in a simplified version with only nearest-neighbor intra-orbital hoppings (all equal for the three 
orbitals).

\subsection{Three-orbital Hubbard model}

\begin{figure}
\includegraphics[width=\columnwidth]{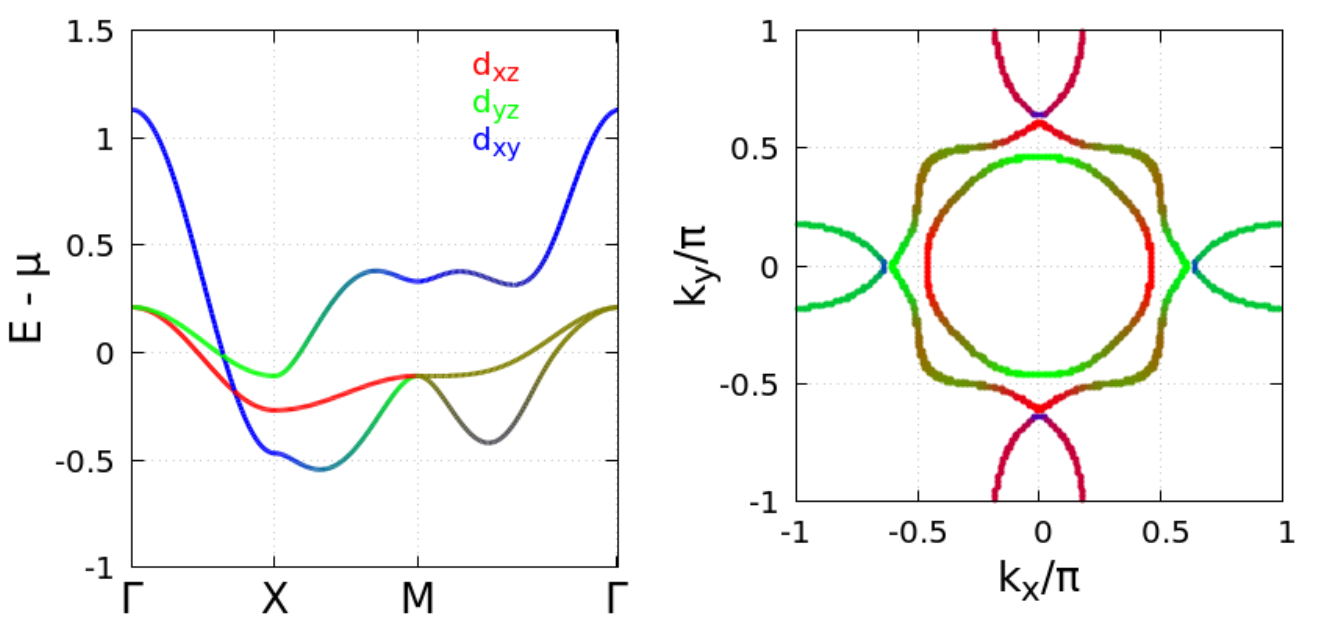}
\caption{\label{fig:FS}
Left panel: Band structure of the tight-binding model of Eqs.~\eqref{eq:TB} along a selected path in the Brillouin zone, connecting the points $\Gamma=(0,0)$, 
$X=(\pi,0)$, and $M=(\pi,\pi)$; the chemical potential at density $n=3+1/3$ is subtracted in order to have the Fermi energy at $E=0$. The color code denotes the 
dominant orbital contribution to each band: $d_{xz}$ (red), $d_{yz}$ (green), and $d_{xy}$ (blue). Right panel: Fermi surface for the electron density $n=3+1/3$.
The color code is the same as in the left panel.} 
\end{figure}

The tight-binding Hamiltonian considered in this work is easily defined in momentum space as:
\begin{equation}\label{eq:TB}
\mathcal{H}_{0} = \sum_k \sum_{\alpha,\beta} \sum_{\sigma} c^\dag_{k,\alpha,\sigma} T_{\alpha,\beta}(k) c^\dagga_{k,\beta,\sigma},
\end{equation}
where $\alpha$ and $\beta$ are orbital indices ($1=xz$, $2=yz$, and $3=xy$), while $c^\dag_{k,\alpha,\sigma}$ ($c^\dagga_{k,\alpha,\sigma}$) is the fermionic 
operator that creates (annihilates) an electron in orbital $\alpha$, with momentum $k$ and spin $\sigma$. The intra- and inter-orbital hoppings are given as: 
\begin{eqnarray}
&&T_{1,1}(k) = 2t_2 \cos{k_x} + 2t_1 \cos{k_y} + 4t_3 \cos{k_x}\cos{k_y}, \nonumber \\
&&T_{2,2}(k) = 2t_1 \cos{k_x} + 2t_2 \cos{k_y} + 4t_3 \cos{k_x}\cos{k_y}, \nonumber \\
&&T_{3,3}(k) = 2t_5 (\cos{k_x} + \cos{k_y}) + 4t_6 \cos{k_x}\cos{k_y} + \epsilon_{xy},  \nonumber \\
&&T_{1,2} = T^{*}_{2,1} = 4t_4 \sin{k_x} \sin{k_y}, \nonumber \\
&&T_{1,3} = T^{*}_{3,1} = 2it_7 \sin{k_x} + 4it_8 \sin{k_x} \cos{k_y}, \nonumber \\
&&T_{2,3} = T^{*}_{3,2} = 2it_7 \sin{k_y} + 4it_8 \sin{k_y} \cos{k_x}. \nonumber
\end{eqnarray}
In this work, we fix the hopping parameters (in units of eV): $t_1 = 0.02$, $t_2 = 0.06$, $t_3 = 0.03$, $t_4 = -0.01$, $t_5 = 0.1$, $t_6 = 0.15$, $t_7 = -0.1$, 
$t_8 = -t_7/2$~\cite{fanfarillo2020}. The orbital $d_{xy}$ has a crystal field $\epsilon_{xy} = 0.2$ (notice that a slight different choice has been proposed in 
Ref.~\cite{daghofer2010}). With the present choice of the hopping parameters, the total bandwidth is $W \approx 1.6$eV. The band structure and the corresponding 
Fermi surface are reported in Fig.~\ref{fig:FS} for density $n=3+1/3$. 

\begin{figure}
\includegraphics[width=\columnwidth]{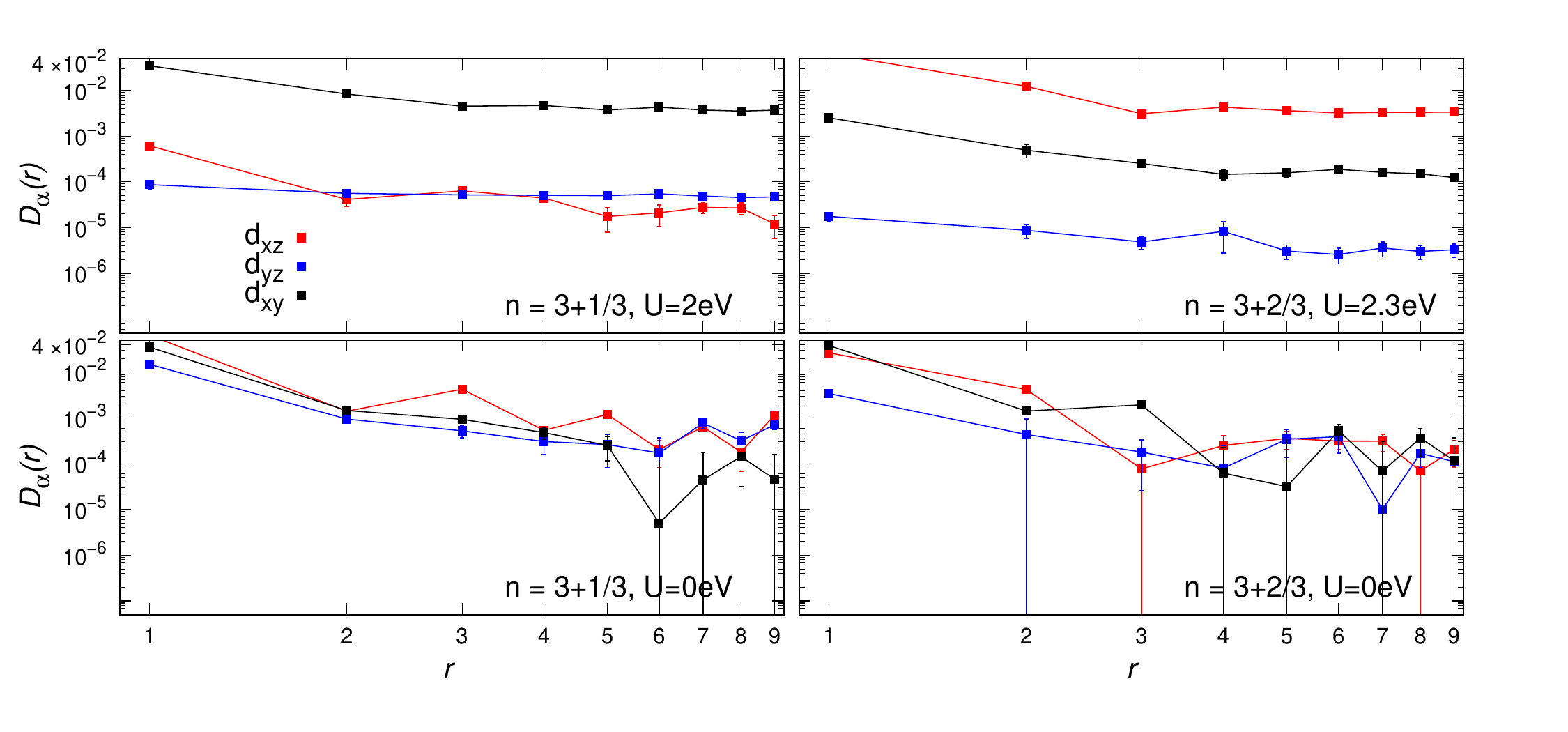}
\caption{\label{fig:Superc_suppl}
Upper panels: $D_{\alpha}(r)$ in all orbitals, for a value of $U$ within the orbital-selective regime. Data are shown at total electronic densities $n=3+1/3$ (left column) 
and $n=3+2/3$ (right column). Lower panels: The same as above, but for $U=0$. Pairing correlations are 
obtained on a $L=18\times18$ cluster at $J/U=0.2$.}
\end{figure}

\begin{figure}
\includegraphics[width=\columnwidth]{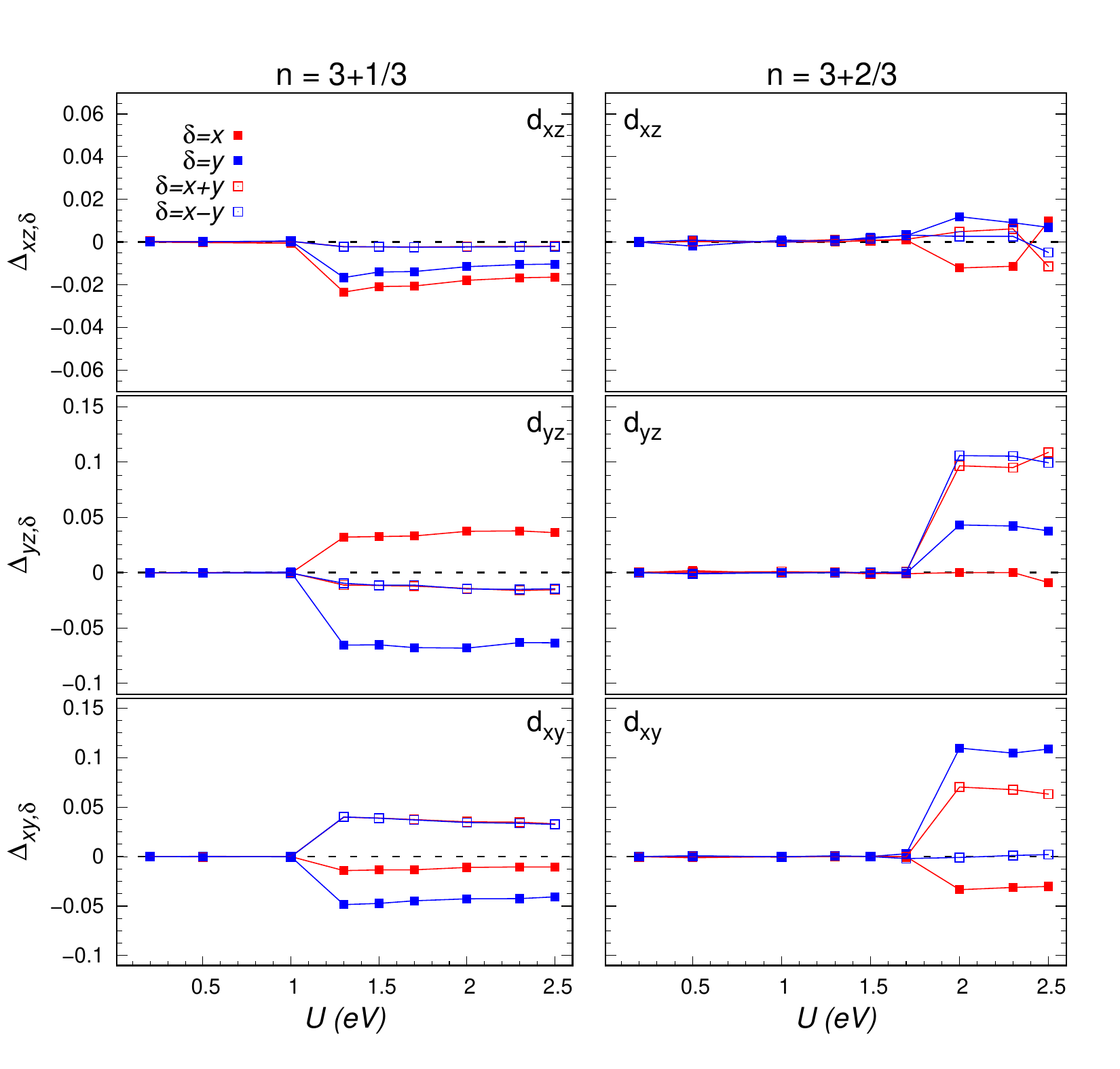}
\caption{\label{fig:BCSr}
Optimal intra-orbital BCS parameters [see  Eq.~(2) of the main text] as a function of $U$. Data are shown at total electronic densities $n=3+1/3$ (left column) and $n=3+2/3$ (right column). BCS parameters are computed with the variational approach on a $L=12\times 12$ lattice at $J/U=0.2$.}
\end{figure}

The interaction terms are defined by the Hubbard-Kanamori Hamiltonian:
\begin{eqnarray}
&&\mathcal{H}_{\rm int} = U \sum_{R} \sum_{\alpha} n_{R,\alpha,\uparrow} n_{R,\alpha,\downarrow} \nonumber \\
&&+U^\prime \sum_R \sum_{\alpha \neq \beta} n_{R,\alpha,\uparrow} n_{R,\beta,\downarrow} \nonumber \\
&&+(U^\prime-J) \sum_R \sum_{\alpha > \beta} \sum_{\sigma}  n_{R,\alpha,\sigma} n_{R,\beta,\sigma} \nonumber \\
&&-J \sum_R \sum_{\alpha \neq \beta} c^\dag_{R,\alpha,\uparrow} c^\dagga_{R,\alpha,\downarrow} c^\dag_{R,\beta,\downarrow} c^\dagga_{R,\beta,\uparrow} \nonumber \\
&&+J \sum_R \sum_{\alpha \neq \beta} c^\dag_{R,\alpha,\uparrow} c^\dag_{R,\alpha,\downarrow} c^\dagga_{R,\beta,\downarrow} c^\dagga_{R,\beta,\uparrow},
\end{eqnarray}
where $n_{R,\alpha,\sigma}=c^\dag_{R,\alpha,\sigma} c^\dagga_{R,\alpha,\sigma}$ is the density operator on site $R$, orbital $\alpha$ and spin $\sigma$; $U$ and 
$U^\prime$ are the intra-orbital and inter-orbital Hubbard interactions, respectively, and $J$ is the Hund's coupling. We assume the system to be rotationally invariant, and thus $U'=U-2J$~\cite{georges2013}.

\subsection{Pairing correlations at $U=0$}

In Fig.~\ref{fig:Superc_suppl}, we report the intra-orbital pairing correlations $D_{\alpha}(r)$ at densities $n=3+1/3$ and $n=3+2/3$, both within the orbital-selective phase and at $U=0$. In the former case, we observe that pairing correlations at large distance are clearly enhanced going from $U=0$ to $U=2$eV. On the contrary, in $d_{xz}$ and $d_{yz}$ orbitals, they are suppressed when electron-electron correlation is switched on. These features clearly indicate that only the $d_{xy}$ orbital becomes superconductive. At density $n=3+2/3$, the situation is completely different. Pairing correlations at large distance are strongly enhanced in the $d_{xz}$ orbital when moving from $U=0$ to $U=2.3$eV, while they remain similar or are even suppressed in the other two orbitals when electron-electron correlation is present. In this case, superconductivity develops in the $d_{xz}$ orbital.

\begin{figure}
\includegraphics[width=\columnwidth]{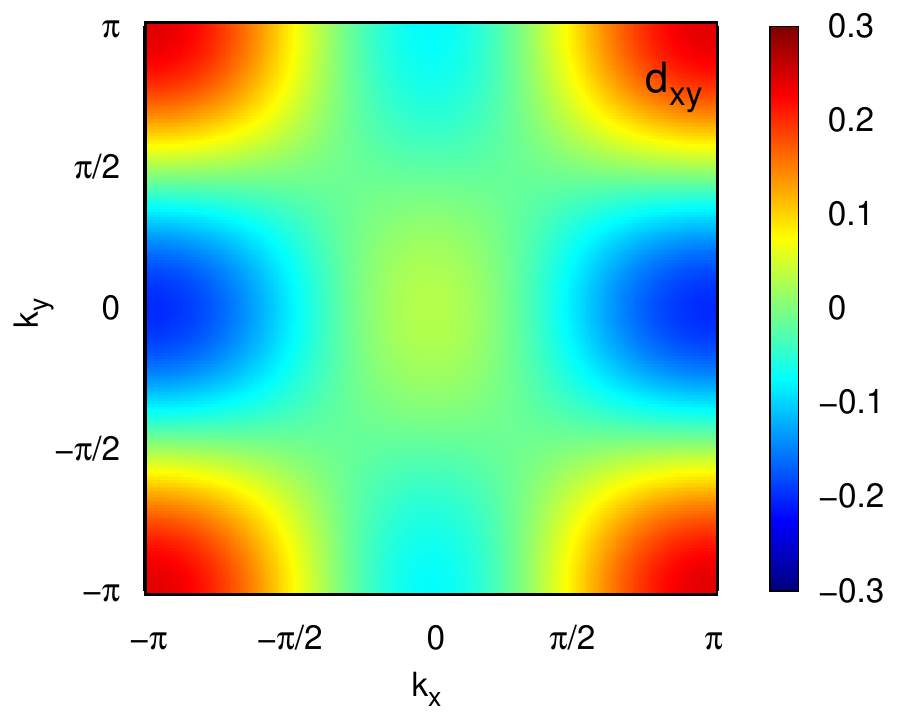}
\caption{\label{fig:BCSk}
Fourier transform $\Delta_{\alpha,k}$ for the $d_{xy}$ orbital, computed for the optimal BCS parameters at density $n=3+1/3$ in the orbital-selective regime at 
$U=2$eV and $J/U=0.2$. Data are reported on the $L=12\times 12$ cluster.}
\end{figure}

\begin{figure}
\includegraphics[width=\columnwidth]{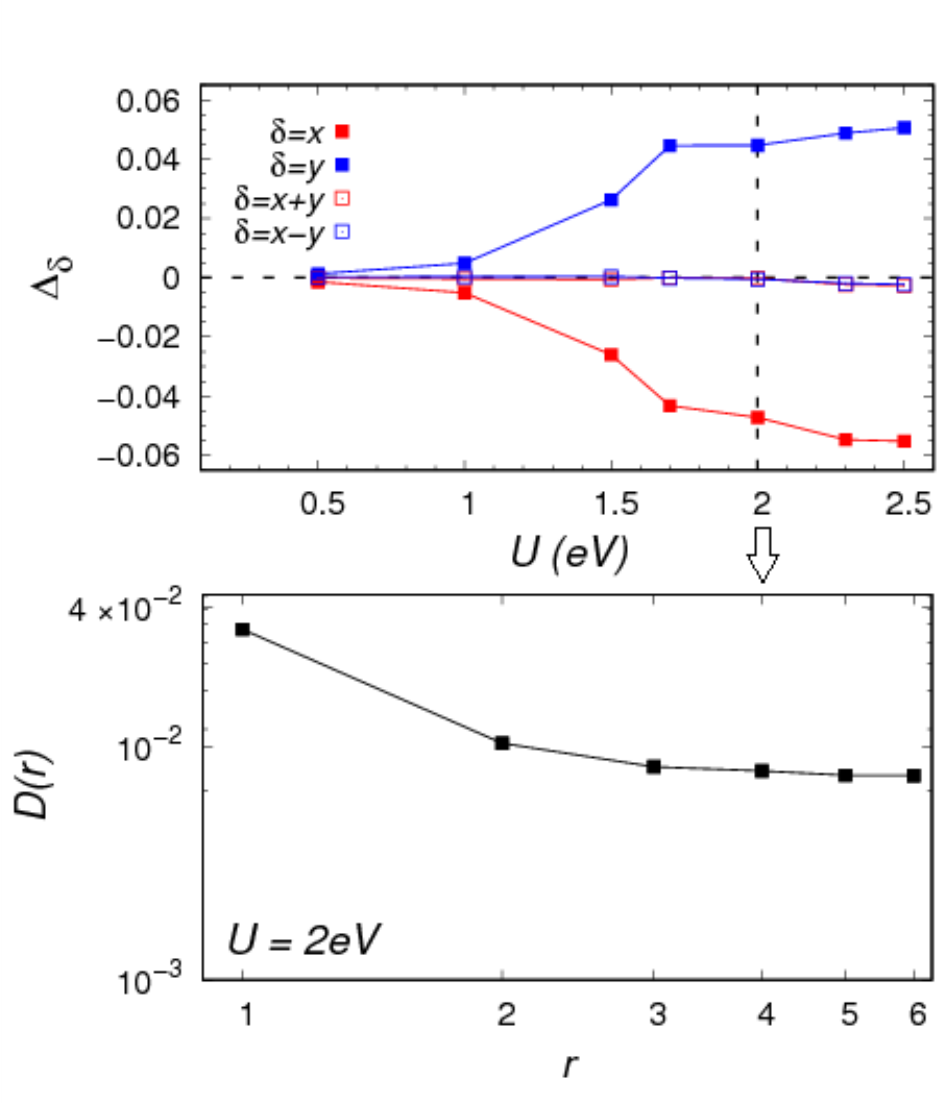}
\caption{\label{fig:equalt}
Upper panel: Optimal intra-orbital BCS parameters (the same in each orbital), as a function of $U$ for $n=3+1/3$, when the tight-binding Hamiltonian contains 
only a nearest-neighbor hopping $t=-0.3$eV, see Eq.~\eqref{eq:onlyt}. Lower panel: Pairing-pairing correlations $D(r)$ [see Eq~\eqref{eq:D_r}] at $U=2$eV, in 
the region where the optimal BCS parameters are finite. Results are obtained with the variational approach on a $12 \times 12$ cluster, for $J/U=0.2$.} 
\end{figure}

\subsection{Optimal BCS parameters}

Within the VMC approach, the presence of superconductivity is
usually associated with the stabilization of BCS parameters in the optimized wave function, see Eq.~(2) in the main text, that defines the uncorrelated state $\ket{\Phi_0}$. Still, pairing in the uncorrelated part of the wave function does not imply superconducting order, since the Jastrow factor may modify or even suppress the 
actual pairing correlations (e.g., in the Mott regime~\cite{capello2005,defranco2018}). 

In Fig.~\ref{fig:BCSr}, we show the optimal (intra-orbital) BCS parameters $\Delta_{\alpha,\delta}$ of Eq.~(2) of the main text, for  $n=3+1/3$ and $n=3+2/3$. In both cases, at sufficiently large values of the Hubbard-$U$ interaction, the orbital selective phase is accompanied by the development of sizable BCS parameters in both $d_{xy}$ and $d_{yz}$ orbitals, while they are relatively small in the $d_{xz}$ one. However, the presence of inter-orbital hoppings in the Hamiltonian, in addition to the Jastrow factors, induces a remarkable difference in the pairing correlations: For $n=3+1/3$, the stabilization of finite BCS parameters in the $d_{xy}$ orbital leads to a true superconducting (long-range) correlation $D_{xy}(r)$, see Fig.~\ref{fig:Superc_suppl}. Instead, the other two orbitals, whose densities are close to half-filling, show a very small signal (in this case, the Jastrow factor gives a substantial suppression of pairing 
correlations). The situation is different for $n=3+2/3$, where 
$n_{xy} \approx 1$ and the pairing correlations $D_{xy}(r)$ are much reduced (again by the Jastrow factor) with respect to the previous case, see Fig.~\ref{fig:Superc_suppl} ($U=2.3$eV). However, surprisingly, the $d_{xz}$ orbital (which is not half-filled) shows sizable pairing correlations, despite having relatively small BCS parameters. This fact is only possible because of the inter-orbital hybridization in the BCS Hamiltonian.

\subsection{Fourier transform of the BCS parameters at density $n=3+1/3$}

The symmetry of the BCS parameters is highlighted in $k$-space, where:
\begin{eqnarray}
&& \Delta_{\alpha,k} =2 \left[ \Delta_{\alpha,x} \cos(k_x) + \Delta_{\alpha,y} \cos(k_y) \right. \nonumber \\ 
&&+ \left. \Delta_{\alpha,x-y} \cos(k_x-k_y) + \Delta_{\alpha,x+y} \cos(k_x+k_y) \right].
\end{eqnarray}
In particular, the results for the $d_{xy}$ orbital (where superconductivity emerges at $n=3+1/3$) are reported in Fig.~\ref{fig:BCSk}, inside the nematic regime 
(for $U=2$eV). Here, $\Delta_{xy,k}$ changes sign between the hole pocket around $\Gamma=(0,0)$ and the electron pockets around the $X=(\pi,0)$ and $Y=(0,\pi)$, 
resembling the $s^{\pm}$ symmetry. The underlying broken symmetry induces a nematic BCS pairing in the $d_{xy}$ orbital, with a clear difference between $X$ and 
$Y$ points. This feature derives from the fact that, in real space, the BCS parameters connecting nearest-neighbor sites are different, with 
$|\Delta_{xy,y}| > |\Delta_{xy,x}|$.

\subsection{Results for the simplified multiorbital model}

Here, we show the results for the simplified tight-binding model that is defined by considering only a nearest-neighbor intra-orbital hopping $t$, that is the 
same for the three orbitals. The hopping matrix in momentum space is then defined as:
\begin{equation}\label{eq:onlyt}
T_{\alpha,\beta}(k) = 2t [\cos{k_x} + \cos{k_y}] \delta_{\alpha,\beta}.
\end{equation}

We report in Fig.~\ref{fig:equalt} the results for the pairing amplitudes $\Delta_{\delta}$ and for the pairing-pairing correlations $D(r)$, defined as:
\begin{equation}\label{eq:D_r}
D(r) = \frac{1}{L} \sum_{R} \langle P^\dagga_{R} P^\dag_{R+rx} \rangle,
\end{equation}
where $P^\dagga_{R} = c^\dagga_{R+y,\downarrow} c^\dagga_{R,\uparrow} - c^\dagga_{R+y,\uparrow} c^\dagga_{R,\downarrow}$. In this case, the orbital index is 
dropped since the results are identical for all the orbitals.

Our results show that this model displays superconductivity for $n=3+1/3$, but with the $d_{ x^2-y^2}$ symmetry, that one obtains also for the one-band Hubbard 
model~\cite{tocchio2016}.

\bibliography{bibliography}

\end{document}